# Nanoelectronics with Two-Dimensional Magnets

Bing Zhao[1*], Roselle Ngaloy[1], Lalit Pandey[1], Himanshu Bangar[1], Divya P. Dubey[1], Saroj P. Dash[1*]

[1]*Department of Microtechnology and Nanoscience, Chalmers University of Technology,*

*SE-41296, Gothenburg, Sweden*

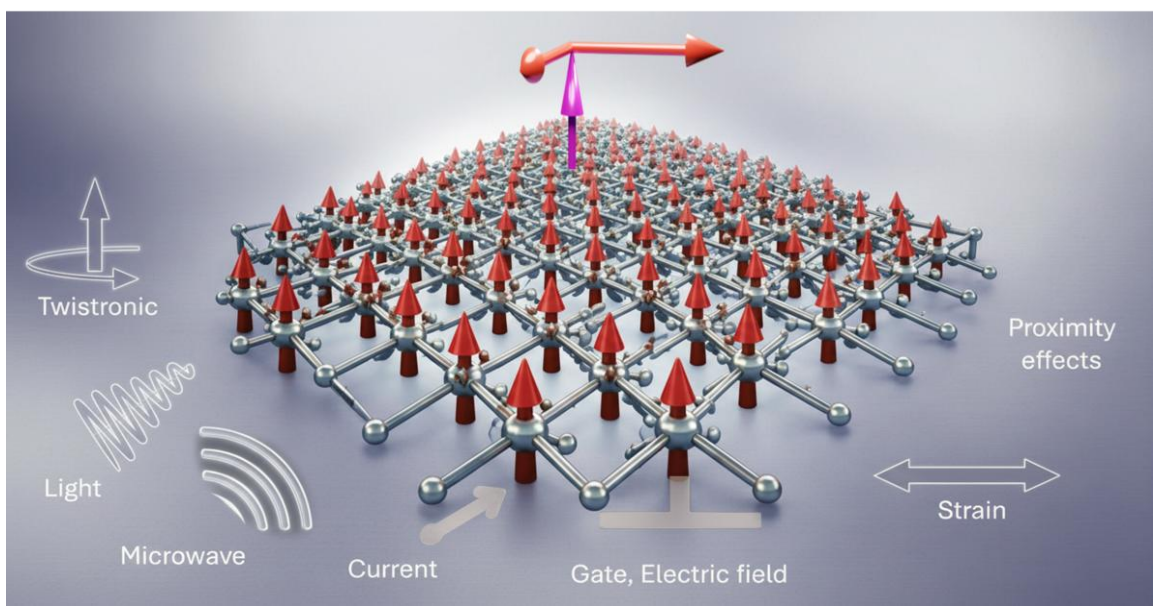

**Abstract:** Two-dimensional (2D) magnets have emerged as a compelling platform for spin-based nanoelectronics, enabling atomic-scale control of magnetic order, interfaces, quantum geometry, and symmetry. Here, we highlight recent advances in 2D ferromagnets, antiferromagnets, altermagnets, and related magnetic phases, emphasizing how enhanced Curie temperatures, perpendicular magnetic anisotropy, and unconventional magnetic orders translate into device-relevant functionality. Spin-dependent transport in vertical magnetic tunnel junctions and lateral spin valves based on 2D heterostructures are discussed, where atomically sharp interfaces enable highly tunable spin injection, propagation, and detection. We further focused on field-free energy–efficient spin–orbit torque magnetization switching in 2D magnetic heterostructures, in which unconventional spin currents originate from an adjacent low–symmetry spin–orbit layer. Microscopic mechanisms involving symmetry breaking, Berry curvature, and orbital angular momentum transport are discussed, along with key challenges, including switching determinism and torque efficiency. Materials and device design strategies targeting neuromorphic, hybrid quantum spintronic, and multifunctional architectures are outlined. Collectively, these developments position 2D magnets as a promising candidate for tunable, energy-efficient integrated spintronic technologies that can harness intertwined spin, charge, orbital, and topological degrees of freedom at the nanoscale.

Corresponding authors: **Saroj P. Dash, Email:** saroj.dash@chalmers.se

**Introduction**

**Spin**, a century after its birth, remains one of the most fundamental degrees of freedom in modern science and technology[1]. Presently, spintronics has emerged as a branch of nanoelectronics that utilizes the electron's spin degree of freedom to write, store, and read information in solid-state devices at room temperature, and hence is suitable for more practical device applications[2,3] (**Fig. 1**). Spintronics has already revolutionized the data storage industry and entered the memory market with non-volatile magnetic random access memory (MRAM) technology. Spintronic components have strong market potential, forecast to grow into a multi-billion-dollar industry over the next decade, because they promise ultra-low-power, high-speed, and non-volatile memory, logic, sensors, and neuromorphic computing technologies.

The core device concepts behind the **rise of spintronics**, such as giant magnetoresistance (GMR)[4,5], tunnel magnetoresistance (TMR)[6,7], and spin transfer torque (STT)[8,9], which have become the cornerstone for efficient generation, manipulation, and detection of spin-polarized current using metallic heterostructures and tunnel junctions. GMR and TMR describe the change in the electrical resistance of a multilayer stack as a result of the relative magnetic or spin orientation of the ferromagnetic layers, while STT is utilized to change this relative orientation by using a spin-polarized current to directly transfer angular momentum and switch the magnetic orientation of the ferromagnetic layer. Recently, spin-orbit torque (SOT) has emerged as an energy-efficient mechanism for driving magnetization dynamics, which generates a transverse spin current from a high-spin-orbit material to manipulate an adjacent ferromagnet[10,11]. Unlike conventional charge-based electronics, spintronics offers nonvolatile operations, leading to minimal standby power and enabling the integration of logic and memory to overcome the von Neumann architecture, thereby achieving faster processing speeds[12].

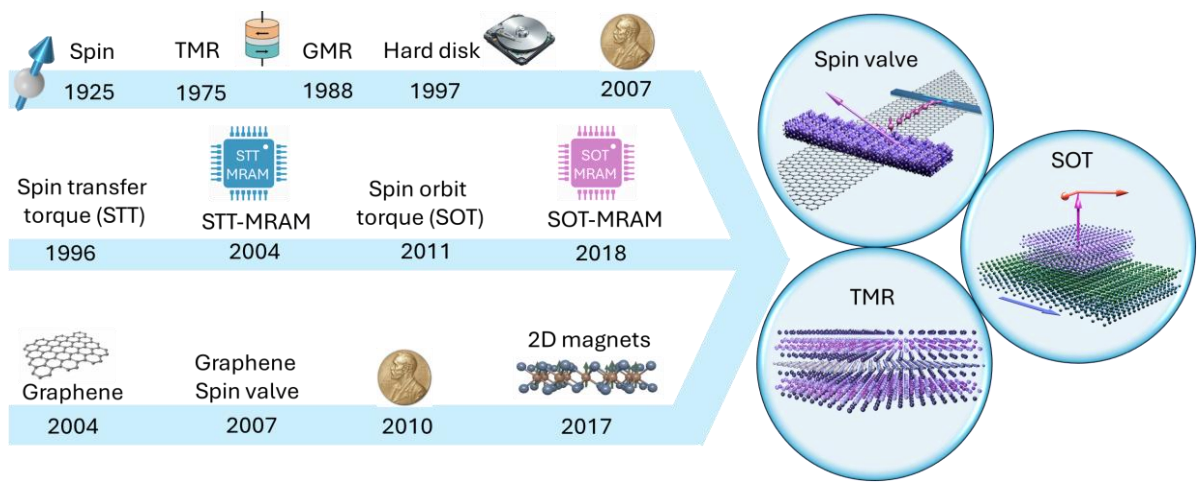

***Figure 1 | Timeline of key milestones in spintronics and the emergence of 2D magnet-based nanoscience and technologies.*** *Historical progression of discoveries and technological breakthroughs in spintronics (first two rows) and 2D materials (third row), from the early conceptualization of electron spin and magnetoresistive effects to the advent of tunnel magnetoresistance (TMR), giant magnetoresistance (GMR), spin valves, spin transfer torque (STT), spin-orbit torques (SOT), and their utilization in magnetic sensors, hard disk drives and magnetic random-access memory (MRAM) technologies. The timeline in the third row highlights the discovery of graphene and 2D magnets, followed by integration into spintronic device platforms. These developments result in 2D materials based spintronic devices, including spin valves, MTJs, and SOT devices. Importantly, Nobel prizes in Physics were awarded in 2007 and 2010 for spintronics and graphene, respectively.*

In more than two decades, **spintronic devices**, particularly memory devices, have matured from a fundamental proof-of-concept to commercially available components. Spintronics offers a range of promising options, including magnetic random-access memories (MRAM)[10,11,13,14], racetrack memory[15], spin logic[16], magnonic devices[17], and neuromorphic computing[18,19]. Despite these advances, several critical challenges remain. One of the fundamental problems in spintronics is the spin conductance mismatch between ferromagnets and nonmagnetic materials, which limits efficient spin injection and detection[20]. Interface engineering has also become a key factor as interdiffusion, interface degradation, and uncontrolled hybridization at the interfaces limit device performance and reproducibility. Moreover, most of the spintronic devices use ultrathin (1.2 nm) ferromagnets such as CoFeB to obtain interfacial perpendicular magnetic anisotropy (PMA)[21]. Although PMA ferromagnets are desired for high-density device integration, they are achieved through complex multilayer interfaces, making them more sensitive to surface and interface degradation issues. Therefore, atomically thin **two-dimensional (2D) magnets** are a promising solution to address these challenges (**Fig. 2**). The use of 2D materials offers several advantages, such as atomically thin thickness, ultra-flat interface, flexibility, and extreme sensitivity to external stimuli. This allows for large gate tunability and strong interface proximity interactions, which can tune and optimize desirable properties and improve device performance.

In this context, **2D magnets** can enable the realization of magnetic tunnel junctions (MTJs) in the atomic thickness limit. Unlike conventional MTJs, which consist of two ferromagnetic electrodes separated by an ultrathin oxide barrier, recent studies have demonstrated that large tunneling magnetoresistance can be achieved using stacked or twisted 2D antiferromagnetic layers sandwiched between metallic electrodes[22]. These systems introduce additional degrees of freedom, such as crystallographic alignment and twist angle, offering new mechanisms for spin-dependent tunneling. Furthermore, conventional spin-orbit materials predominantly generate in-plane SOT components, which generally require an external magnetic field to deterministically switch perpendicularly magnetized ferromagnets. Van der Waals (vdW) heterostructures composed of 2D spin-orbit materials and 2D ferromagnets provide an elegant solution to this limitation. Several low-symmetry 2D spin-orbit materials, such as $WTe_2$, $NbSe_2$ and $TaIrTe_4$ exhibit unconventional spin Hall effects that generate current-induced out-of-plane spin polarizations[23–27]. These out-of-plane SOT components enable field-free magnetization switching in 2D ferromagnets at remarkably low current densities[23,28–31].

Here, we provide a comprehensive overview of recent progress in 2D magnets and their prospects in spin-based nanoelectronics. As the number of new phenomena and publications on 2D magnets continues to increase, this review provides a timely update on materials, devices, and future perspectives. We first outline the materials landscape of 2D ferromagnets (FM), antiferromagnets (AFM), altermagnets, and other related magnetic phases, emphasizing the microscopic mechanisms that stabilize magnetic order in the 2D limit. We then focus on spin-dependent transport phenomena enabled by 2D heterostructures, including magnetic tunnel junctions and spin valve devices, where atomically sharp interfaces, twist-angle control, and proximity effects offer new degrees of freedom for spin injection, transport, and detection. Attention is devoted to current-driven magnetization dynamics in 2D systems, covering both all vdW heterostructure-based spin-orbit torques and the emerging self-induced spin-orbit torques in metallic vdW ferromagnets. Finally, we discuss how interfacial engineering, orbital and symmetry effects, and reduced dimensionality enable novel device concepts ranging from field-free switching and multistate memory to neuromorphic, magnonic, and hybrid quantum spintronic architectures. Together, this perspective highlights how vdW magnets provide a unified platform for exploring emergent spin–orbit physics and for advancing next-generation spintronic technologies beyond conventional material and device paradigms.

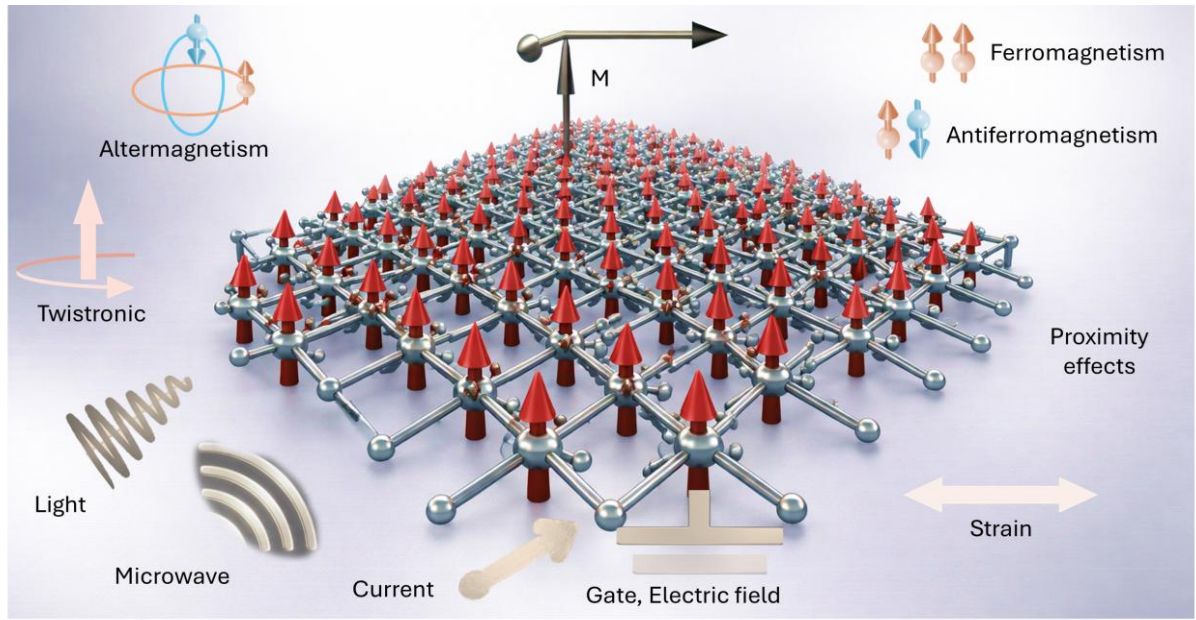

*Figure 2 | Tunable two-dimensional magnets.* Schematic illustration of a 2D magnet with out-of-plane magnetic moments M and its dynamical modulation. The magnetic order and collective excitations can be engineered through multiple external perturbations, including electric gating/electric fields, applied current, strain or pressure, and electromagnetic radiation spanning microwave to optical frequencies. Additional control arises from twist-induced moiré superlattices (twistronics) and proximity effects from adjacent materials, enabling modification of spin-orbit and magnetic exchange interactions, anisotropy, and spin textures.

**Magnetism in the two-dimensional limit**

The isolation of graphene[32] initialized the exploration of atomically thin materials with properties that are distinct compared to their bulk counterparts. Until 2016, magnetic order in purely 2D crystals remained elusive, as suggested by Mermin-Wagner theorem[33], which forbids long-range magnetic order in an isotropic 2D Heisenberg system at finite temperatures. This barrier was overcome by the discovery of intrinsic ferromagnetism in monolayer and few-layer vdW crystals, which was first demonstrated in $CrI_3$[34] and independently in $Cr_2Ge_2Te_6$[35]. In these seminal works, they have reported clear hysteresis loops and layer-dependent remanent magnetization down to the monolayer limit, firmly establishing a magnetic class of 2D layered materials stabilized by magnetocrystalline anisotropy. Since those initial reports, the library of 2D magnets has grown rapidly to include ferromagnets, antiferromagnets, ferrimagnets, multiferroics, and materials exhibiting more exotic spin textures[36–38]. These include metallic $Fe_3GeTe_2$[39], semiconducting $VI_3$[40], insulating $MnPS_3$[41], topological magnetic insulator $MnBi_2Te_4$[42], and multiferroic $CuCrP_2S_6$[43], among others, which highlights the diversity in magnetic anisotropy, magnetic ground state, electronic structure, and spin-orbit coupling strength.

Moreover, 2D magnets exhibit thickness-dependent magnetism, where their magnetic ordering can drastically change in few-layer forms, leading to fascinating interlayer effects[44]. Pressure, strain, electrostatic gating, and twisting can also tune the magnetic order of these materials (**Fig. 2**), effectively controlling phase transitions between paramagnetic, ferromagnetic, antiferromagnetic, and altermagnetic states[45,46,55,56,47–54]. Beyond these intrinsic 2D magnets, heterostructures that combine them with graphene, transition metal dichalcogenides (TMDs), topological insulators (TIs), and superconductors have revealed proximity-induced magnetic effects, spin filtering, and even novel quantum states that could be harnessed for future quantum technologies, for sensing, memory and computing[45,57–61].

**Magnetic order** originates from the antisymmetric nature of the electron wavefunction and is determined by exchange interactions between neighboring spins. Depending on the degree of electronic localization and orbital overlap, magnetic coupling can arise from direct exchange[62], super exchange mediated by nonmagnetic ligands[63], or indirect exchange via itinerant carriers, as described by the Ruderman–Kittel–Kasuya–Yasuda (RKKY) mechanism[64–66]. These interactions underpin the diverse magnetic ground states observed in bulk materials, ranging from simple ferromagnets to noncollinear and spiral spin textures. Motivated by the robust magnetism of elemental ferromagnets and their alloys, early efforts to realize low-dimensional magnetism focused on epitaxial thin films. However, such systems often suffer from surface roughness, interfacial hybridization, and limited reproducibility, obscuring intrinsic two-dimensional (2D) magnetic behavior[67–69]. In the strictly two-dimensional limit, long-range magnetic order faces a fundamental constraint imposed by the Mermin–Wagner theorem, which prohibits finite-temperature magnetic order in isotropic systems with continuous spin symmetry and short-range interactions[33]. Thermal population of long-wavelength spin-wave excitations lead to divergent fluctuations that destabilize magnetic order in low-dimensional isotropic spin systems. However, magnetic anisotropy, typically arising from spin–orbit coupling, circumvents this restriction by breaking continuous spin rotational symmetry and opening a gap in the magnon spectrum, thereby suppressing low-energy fluctuations. This principle underlies the stabilization of intrinsic magnetism in atomically thin 2D vdW materials, where reduced dimensionality coexists with sizable magnetic anisotropy.

The **first experimental demonstrations** of intrinsic 2D magnetism in vdW crystals, notably $CrI_3$ and $Cr_2Ge_2Te_6$[34,35], established this paradigm (**Fig. 3a**). In $CrI_3$, strong out-of-plane Ising anisotropy originating from heavy iodine atoms enables magnetic order to persist down to the monolayer limit[70], as revealed by magneto-optical Kerr effect measurements[34]. The layered nature of $CrI_3$ further gives rise to pronounced thickness-dependent magnetic behavior, including odd–even effects arising from ferromagnetic intralayer and antiferromagnetic interlayer coupling. In contrast, $Cr_2Ge_2Te_6$ represents a rare realization of a 2D Heisenberg ferromagnet with weak magnetic anisotropy, where long-range order can be stabilized only under an external magnetic field[35]. Together, these materials highlight the central role of anisotropy in defining magnetic stability in two dimensions. Since these initial discoveries, the library of vdW magnets has expanded rapidly to include ferromagnetic[39], antiferromagnetic, ferrimagnetic and multiferroic systems spanning insulating, semiconducting, and metallic electronic structures[34,35,78–87,39,88–97,71,98,99,72–77]. Notably, metallic vdW ferromagnets such as $Fe_3GeTe_2$[47] and $Fe_3GaTe_2$[73] exhibit much higher Curie temperatures, with magnetic order persisting at or above room temperature in few-layer form. These materials mark a critical milestone toward spintronic devices and illustrate how itinerant magnetism and enhanced spin–orbit coupling can coexist in the 2D limit. A broader overview of known 2D magnetic materials, their transition temperatures, and magnetic ground states is summarized in **Fig. 3b**, underscoring the rapidly diversifying landscape of low-dimensional magnetism.

Beyond conventional collinear magnetic order, reduced dimensionality and symmetry in vdW systems enable a range of emergent magnetic phases (**Fig. 3c**). One prominent example is intrinsic **topological magnetism,** realized in layered materials such as $MnBi_2Te_4$[100]. In this compound, ferromagnetic intralayer exchange combined with antiferromagnetic interlayer coupling gives rise to an antiferromagnetic topological insulator state. The vdW layered structure allows thickness and symmetry-controlled access to distinct topological phases, including axion insulator states[100], without relying on magnetic doping or proximity effects. More broadly, the coexistence of magnetism, strong spin–orbit coupling, and reduced crystal symmetry enables symmetry-engineered transport phenomena such as nonlinear Hall responses[101] and antiferromagnetic diode effects[102], demonstrating how topology and magnetism become intrinsically intertwined in two dimensions.

These symmetry-driven phenomena also favor the emergence of chiral real-space spin textures, **skyrmions**, including Néel- and Bloch-type domain walls, skyrmions, and related topological spin configurations[103]. In atomically thin vdW heterostructures, interfacial inversion-symmetry breaking and strong spin–orbit coupling generate sizable Dzyaloshinskii–Moriya interactions (DMI), enabling magnetic field-assistant stabilization of Néel skyrmions with characteristic diameters of 10–200 nm[104]. To be noted, bulk vdW magnetic materials with a noncentrosymmetric structure[105] or symmetry breaking by vacancies or interlayer occupation of atom, stacking faults, or strain can locally promote DMI and support topological spin configurations up to room temperature without field assistance[106–109]. Metallic vdW ferromagnets such as $Fe_3GeTe_2$, $Fe_3GaTe_2$, also exhibit chiral domain walls[110,111]. These nanoscale textures can be driven efficiently by spin–orbit torques, with critical current densities typically ~ $10^7$ A cm$^{-2}$ and domain-wall velocities reaching ~10 m s$^{-1}$, much lower than that of conventional heavy-metal/ferromagnet stacks[110,111]. Domain wall velocities in current reports remain modest, highlighting ongoing materials and interface optimization challenges.

Another recently recognized class of magnetic order enabled by symmetry principles is **altermagnetism**[112–114], characterized by zero net magnetization yet momentum-dependent spin-split electronic bands. In the 2D limit, altermagnets can offer a platform where spin, valley, and lattice degrees of freedom can become strongly coupled, giving rise to topological band structures without macroscopic magnetization[115,116]. However, 2D altermagnets have not been realized experimentally yet. Theoretical studies further predict that 2D altermagnets can host complex noncollinear spin textures, including altermagnetic skyrmions exhibiting anisotropic transport responses[117]. These findings broaden the taxonomy of magnetic ground states accessible in two dimensions and challenge conventional distinctions between ferromagnetic and antiferromagnetic order. Structural degrees of freedom provide an additional route to engineering magnetism in the 2D limit. Twisted vdW heterostructures give rise to moiré superlattices, where spatially modulated interlayer coupling leads to emergent electronic and magnetic behavior absent in the parent layers. In twisted bilayers of 2D magnets, moiré magnetism has been predicted and observed to stabilize noncollinear spin textures, merons, and antimerons[118,119], and mixed ferromagnetic–antiferromagnetic domain states[120]. These systems demonstrate that exchange interactions themselves become programmable in the 2D limit, offering a new mechanism for realizing complex magnetic order.

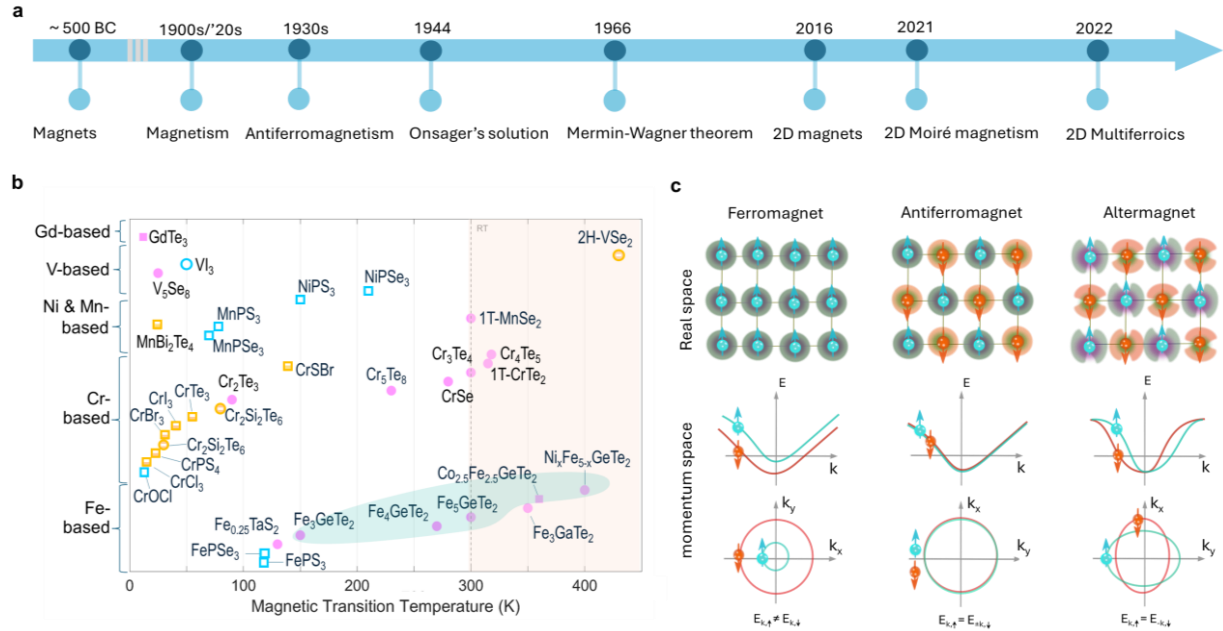

*Figure 3 | Timeline of progress and emerging 2D magnets. **a.** Key milestones in the discovery of magnetic materials and phenomena, including the theory of Onsager's solution and Mermin-Wagner theorem for 2D magnets and other novel magnetic phases. **b.** Library of van der Waals magnets and their reported Curie $T_C$ or Neel $T_N$ temperatures. These include ferromagnets (circles) and antiferromagnets (squares) with different electrical properties: insulating (cyan, empty symbols), semiconducting (yellow, half-filled symbols), metallic (magenta, filled symbols). Materials with the same basic magnetic elements are clubbed together. Data points are obtained from references [34,35,77–86,39,87–96,42,97–99,121,71–76]. **c.** Physical concepts of magnetic ground states: ferromagnetic, antiferromagnetic, and altermagnetic states in real and momentum (k) spaces. In real space, (anti)ferromagnets display isotropic spin density, while it shows anisotropic spin density in altermagnet. In K-space, the energy bands show different splitting for the different magnetic ground states. From a symmetry perspective, ferromagnets and antiferromagnets have inversion symmetry, whereas time reversal symmetry is preserved in antiferromagnets and altermagnets.*

An emerging aspect of 2D magnetism is the role of **orbital magnetization**. Unlike spin magnetization, orbital magnetization originates from the self-rotation of Bloch wave packets and circulating electronic currents[122] and is intimately connected to Berry curvature and band topology[123]. In few-layer vdW systems, reduced dimensionality, strong spin–orbit coupling, and broken inversion symmetry can significantly enhance orbital effects, allowing orbital contributions to compete or even exceed those from spin. This behavior is evident in intrinsic 2D ferromagnets such as $CrI_3$, where orbital magnetism plays a crucial role in determining magnetic anisotropy[124]. Orbital magnetization is also central to topological vdW materials, like $MnBi_2Te_4$, where Berry-curvature-induced orbital moments give rise to unconventional magnetotransport and strongly influence topological surface states[125]. Furthermore, in moiré and twisted vdW heterostructures[126], such as twisted bilayer graphene[127], the flat electronic bands and enhanced quantum geometry generate large orbital magnetic moments, leading to correlated and orbital-dominated magnetic states. The strong sensitivity of orbital magnetization to crystal symmetry, stacking configuration, strain, and electrostatic gating provides powerful and unique control mechanisms for tailoring magnetic responses in atomically thin systems.

Finally, reduced dimensionality also enables strong coupling between magnetic and electric order parameters, offering new opportunities for multifunctional nanoelectronics devices. Engineered vdW heterostructures that combine 2D ferroelectrics and 2D magnets have demonstrated robust magnetoelectric coupling mediated by charge transfer, exchange proximity, strain, and interlayer sliding[128–130]. These interfacial mechanisms allow electric-field control of magnetic anisotropy, exchange interactions, and tunneling spin polarization within atomically thin architectures. Recent observations of spin-driven ferroelectricity (FE) and magnetic-order-induced polarization in atomically thin vdW compounds further indicate that intrinsic **multiferroicity** can emerge from reduced symmetry and competing interactions in the 2D limit[131–133]. In parallel, atomic-scale studies of vdW ferroelectrics such as SnSe have revealed multiple switching pathways[134], while FE and FM heterostructures such as $CuCrP_2S_6$/$Fe_3GeTe_2$ have demonstrated nonvolatile voltage control of magnetism with exceptionally large tunneling electroresistance ratios exceeding $10^6$ percent through interfacial magnetoelectric coupling[128]. These advances highlight the practical advantages of multiferroicity for low-power spintronic applications. Collectively, they establish vdW magnetic heterostructures as a versatile platform in which dimensional confinement, symmetry engineering, and interfacial design can be harnessed to reshape magnetic order and its coupling to electronic, orbital, and lattice degrees of freedom.

**Magnetic tunnel junctions and Spin valve devices with van der Waals magnets**

Magnetoresistance phenomena represent one of the most important spintronic effects in different device architectures, exploiting the giant magnetoresistance (GMR)[4,5] and tunneling magnetoresistance (TMR)[6,7] effects to achieve magnetic field-dependent change in electrical resistance (**Fig. 4**). Spin valves are typically constructed from ferromagnetic (FM) electrodes separated by either a metallic or insulating spacer, enabling controlled spin-dependent electron transport. By contrast, spin valves built from 2D magnets open a new avenue for device engineering: the atomically thin nature of layered materials, combined with their atomically sharp van der Waals interfaces, offers cleaner and more tunable junctions. This eliminates issues of lattice mismatch and interfacial disorder that commonly limit traditional heterostructures. Magnetic tunnel junctions (MTJs) or vertical tunneling spin valves are devices with a ferromagnet and insulating spacer stacked vertically. These devices rely on the change in tunneling resistance as the magnetization of the FM electrode changes. The tunneling current is proportional to the electron density of states (DOS) of the electrodes, which in ferromagnets is intrinsically spin split, meaning that the majority (spin-up) and minority (spin-down) bands exhibit different DOS profiles depending on the magnetization direction. As a result, the junction resistance varies with the alignment of the FM layers. When the magnetizations are parallel, spin-polarized electrons tunnel efficiently, yielding the lowest resistance state ($R_P$). Conversely, when the magnetizations are antiparallel, spin filtering suppresses tunneling, leading to the highest resistance state ($R_{AP}$). The relative contrast between these two states defines the tunneling magnetoresistance (TMR) ratio, a key performance metric for MTJs: *TMR = ($R_{AP}$ – $R_P$)/$R_P$ × 100%*. A tunneling spin valve device was demonstrated using few nanometer thick $Fe_3GeTe_2$ ferromagnetic electrodes separated by an atomically thin hexagonal boron nitride (hBN) as a tunnel barrier[135]. The device exhibited a tunneling magnetoresistance (TMR) ratio as high as 160%, corresponding to a spin polarization of 0.66 for the $Fe_3GeTe_2$ electrodes. This enhanced performance, surpassing conventional magnetic tunnel junctions, was attributed to the atomically sharp and clean interfaces formed between the electrodes and the hBN barrier. Replacing hBN with graphite yielded $Fe_3GeTe_2$/graphite/$Fe_3GeTe_2$ heterostructures exhibiting antisymmetric MR with an intermediate resistance state, explained by Rashba SOC–induced spin–momentum locking at the interfaces[136]. Depending on the relative alignment between spin polarization and magnetization at the two interfaces, low, intermediate, or high MR states were obtained. These results highlight the ability of vdW heterostructures to uncover unconventional spin transport phenomena. Importantly, room-temperature operation with TMR ratios of 50% has been achieved in $Fe_3GaTe_2$/$WSe_2$/$Fe_3GaTe_2$ junctions[137], demonstrating the practical viability of 2D MTJs for memory applications. Importantly, it was demonstrated that magnetic tunnel junctions employing a magnetic insulator as the tunnel barrier can also yield giant tunneling magnetoresistance, reaching values as high as 19,000%[44]. This behavior arises from the spin-filter effect, where tunneling through a magnetic insulator produces spin-dependent barrier heights, leading to different tunneling rates for spin-up and spin-down electrons. Consequently, the observed magnetoconductance strongly reflects the exponential sensitivity of tunneling current to the barrier's electronic structure. Another notable example is the graphite/$CrI_3$/graphite heterostructure[138], which acts as a multilayer spin-filter device due to the interlayer antiferromagnetic ordering and metamagnetic transitions in $CrI_3$. Beyond enabling enhanced device performance, such structures also provide a powerful platform to probe the intrinsic magnetic properties of 2D magnets, offering an electrical alternative to conventional microscopy techniques such as magneto optical Keer effect (MOKE) and Lorentz transmission electron microscopy (LTEM). To be noted, a twist-assisted all-antiferromagnetic tunnel junction (AFMTJ) in the atomic limit, demonstrated using bilayers of 2D antiferromagnet CrSBr, achieves a nonvolatile TMR ratio exceeding 700 % at zero field[22]. This breakthrough stems from the accumulative coherent tunneling across individual CrSBr monolayers, which enables atomic-scale magnetic memory with high stability and sensitivity to interlayer twist angles.

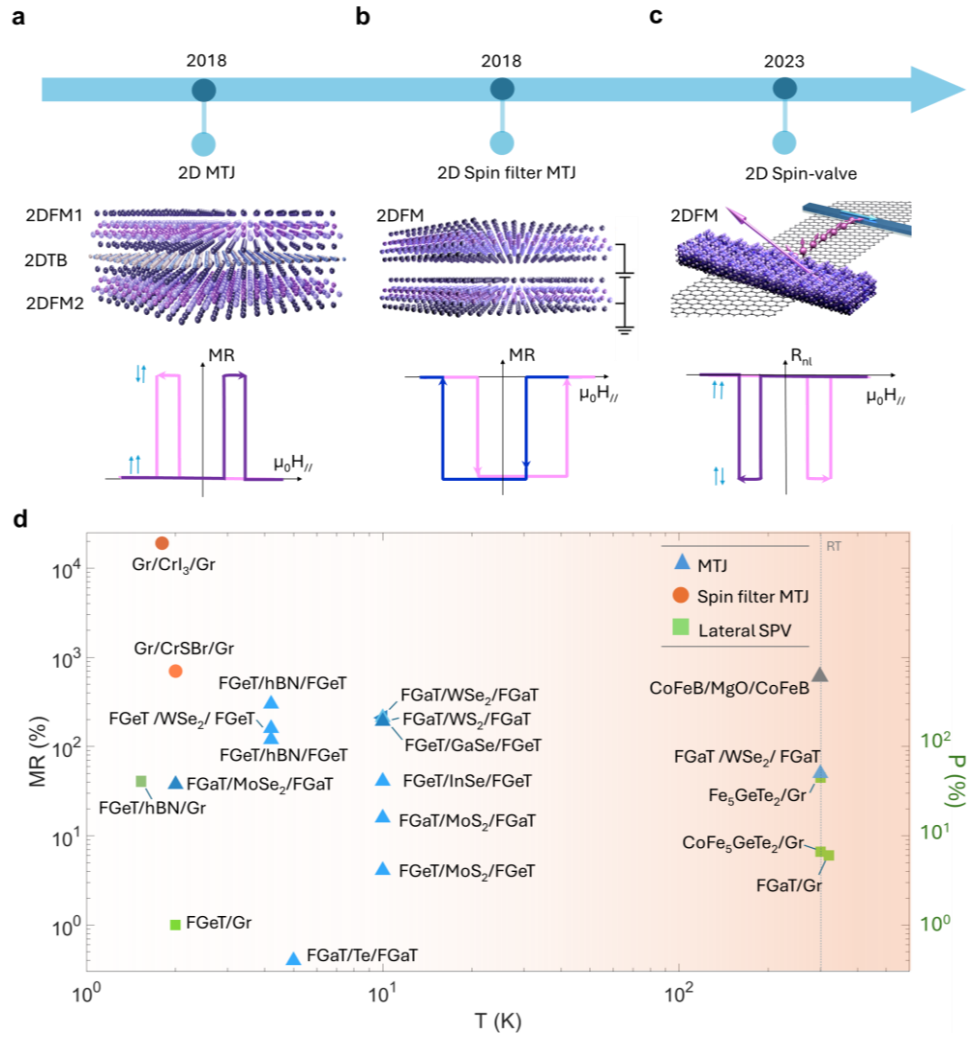

*Figure 4 | **2D magnet-based magnetic tunnel junction and spin valve devices**. **a.** Schematics of 2D MTJ with 2DFM1/2DTB/2DFM2 structure and its representative MR signal vs. external field ($\mu_0H_{\parallel}$). 2DFMs correspond to 2D magnets and 2DTB corresponds to 2D tunnel barrier, such as h-BN. **b.** Schematics of a 2D spin filter MTJ with a bilayer 2DFM structure and its signal MR change with external field ($\mu_0H_{\parallel}$). **c.** Lateral spin valve device with 2DFM as a spin source to induce a non-equilibrium spin accumulation in the graphene channel, which is consequently transported and detected by another FM. The bottom panel shows the typical nonlocal spin valve signal $R_{nl}$ with a magnetic field. The timeline shows the first demonstration of such devices. **d.** Summary of the MR ratio of magnetic tunnel junctions, spin filter magnetic tunnel junctions as a function of temperature, with the industrial MTJ device CoFeB/MgO/CoFeB as reference. Lateral SPV presents spin polarization (P) vs. temperature (T). Abbreviations used are: FGeT ($Fe_3GeTe_2$), FGaT ($Fe_3GaTe_2$), CoFeB ($Co_{40}Fe_{40}B_{20}$). The data were obtained from references[23,44,145–150,135,137,139–144].*

For lateral spin valve devices based on graphene transport channels, the first demonstration of spin injection and detection using a vdW magnet was achieved with $Fe_5GeTe_2$/graphene heterostructures at room-temperature[140]. Efficient spin injections were observed, with a negative spin polarization of 45% in $Fe_5GeTe_2$. This large interfacial polarization arises not only from the high magnetization saturation of $Fe_5GeTe_2$ but also from the atomically sharp interface and vdW gap, with enhanced tunneling barrier properties compared to conventional thin films. More recently, all-vdW heterostructure lateral spin valve devices using $Fe_3GaTe_2$ were reported, showing nonlocal spin valve signals persisting up to room temperature[139]. Together, these advances highlight the potential of vdW ferromagnet/graphene interfaces for efficient spin injection and detection in a non-local spin-valve device. Moreover, the sensitivity of spin valves and nonlocal measurement geometries at the vdW ferromagnet/graphene interfaces opens new possibilities for an all-electrical probe of spin textures in conventional and vdW magnets[151,152].

**Spin-orbit torque and magnetization dynamics in van der Waals magnet devices**

**Spin-orbit torque** is a promising approach for energy-efficient magnetization switching in next-generation spintronic devices. The discovery of 2D magnetic materials has pushed forward research in SOT devices, providing a platform for exploring novel spin-orbit phenomena. This section examines the current state of spin-orbit torque research in 2D magnetic systems, as well as the emerging self-induced spin-orbit torque. For next-generation nonvolatile, magnetic memory devices, SOT has been established as an improvement over spin transfer torque (STT) based technologies. For both SOT and STT, magnetization is switched using electrical current-induced torques, which is attractive for integration in electronic devices. STT has already been successfully utilized in magnetic random-access memory (STT-MRAM)[153]. However, several problems are limiting its performance and device integration: (1) high switching current density, (2) slow switching time in a few nanoseconds, and (3) device instability due to coupled read and write paths[154]. In contrast, SOT-based devices are implemented in a three-terminal magnetic tunnel junction (MTJ) design, where a high in-plane pulse current is introduced to switch the magnetization and a low vertical current is used to detect the magnetization state, separating the read and write paths. The separation of read and write paths significantly improves device endurance and reliability[11,155] (**Fig. 5a**).

For SOT devices in a simplified structure, two layers are needed to build a laboratory test device. First is a high spin-orbit coupling (SOC) material as a spin source layer, which converts charge current to spin current, via spin Hall effect (SHE) or Rashba-Edelstein effect (REE)[156]. Conventional spin sources include heavy metals (HM) such as Pt, Ta, W, and their alloys - materials exhibiting giant spin Hall effect. In addition to this, recent works have been directed towards exploring 2D van der Waals materials with high SOC and topological spin texture, such as topological semimetals and insulators, and transition metal dichalcogenides (TMDs), where charge-to-spin conversion could originate from bulk, surface, or a combination of both[157] (**Fig. 5b**). The second layer is the SOT device with a magnetic layer, which provides the switchable magnetization states corresponding to the memory states. In conventional magnet/HM heterostructures, a charge current injected along the x-axis into a high SOC material generates a transverse spin current that propagates along the out-of-plane direction (z-axis). This spin current diffuses across the interface into the adjacent magnetic layer, where the transfer of angular momentum exerts spin-orbit torques that can switch the magnetization state. In materials characterized by conventional charge-to-spin conversion, the resulting spin polarization is typically oriented along the y-axis, giving rise to in-plane damping-like and field-like torque components. Such torques can efficiently switch in-plane magnets. However, for magnets with PMA, which are desired for fast (sub-nanosecond) switching and high-density devices integration (**Fig. 5c - 5d**), the symmetry of the conventional spin-orbit torque does not inherently distinguish between up and down magnetization states. As a result, deterministic switching of a perpendicular magnet (PMA) generally requires an additional symmetry-breaking mechanism. In laboratory-level practice, a small in-plane magnetic field is applied to break this symmetry and enable controlled switching of PMA magnets. Field-assisted switching has been successfully demonstrated in $Cr_2Ge_2Te_6/Ta$[158], $Cr_2Ge_2Te_6/Pt$[159], $Fe_3GeTe_2/Pt$[160], $Fe_3GeTe_2/Bi_2Te_3$[59], $Fe_3GaTe_2/Pt$[161] (**Fig. 5e**). As field-assisted switching adds further complications to device design, efforts have been directed towards circumventing this limitation. Among the solutions is the utilization of low symmetry materials with unconventional charge-to-spin conversion, such as $WTe_2$ and $TaIrTe_4$ to obtain out-of-plane torques[24,25,162] (**Fig. 5f**). Successful field-free switching has been demonstrated in $Fe_3GeTe_2/WTe_2$[30], $Fe_3GaTe_2/WTe_2$[31], and $Fe_3GaTe_2/TaIrTe_4$[23,28,29]. Another way to realize deterministic switching is through exchange bias in AFM/FM heterostructures, such as demonstrated in Co-doped $Fe_5GeTe_2/Pt$[163] and $CrSBr/Fe_3GeTe_2/Pt$[164], as well as employing intrinsically canted magnets like $Fe_5GeTe_2$[165]. With exchange bias or canted magnetism, the magnetic inversion symmetry is broken, which normally requires an external magnetic field for deterministic magnetization switching.

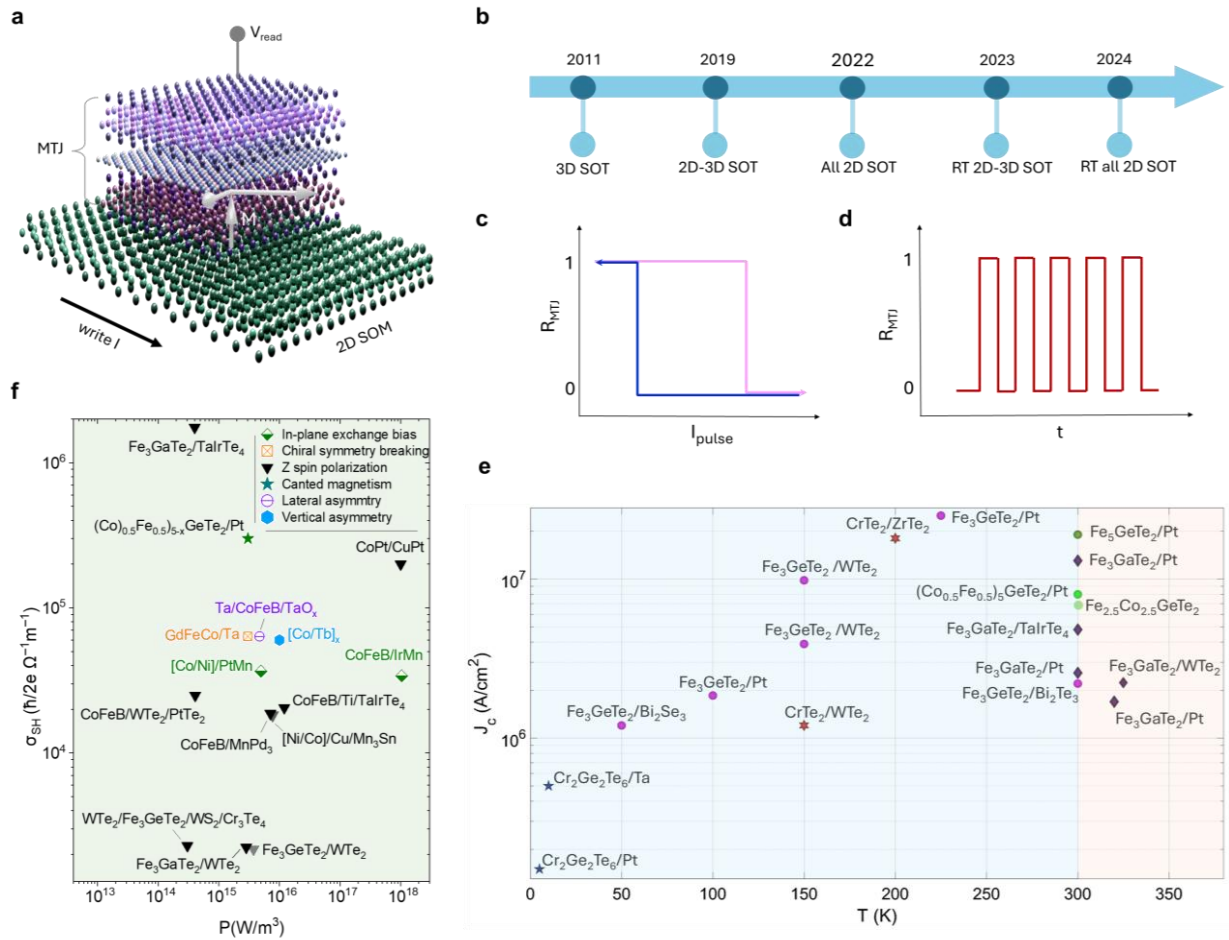

*Figure 5 | Van der Waals magnet-based spin-orbit torque devices.* **a.** *An all van der Waals heterostructure of a nonvolatile spintronic memory component: This integrates 2D spin-orbit material (SOM) and a 2D magnet for realizing energy-efficient and field-free magnetization switching. An MTJ on top of the SOT heterostructure can be used for the electrical read-out of the magnetization switching.* **b.** *Timeline of the key SOT device developments. 3D SOT used all conventional metallic layers; 2D-3D SOT indicates SOT devices with 2D FM and 3D spin-orbital material. All 2D corresponds to the use of both vdW spin-orbit material and vdW ferromagnet. RT corresponds to room-temperature operation.* **c.** *Schematic representation of readout of magnetization switching signal $R_{MTJ}$ using an MTJ with pulse current $I_{pulse}$.* **d.** *Schematics of switching signal $R_{MTJ}$ as a function of time t to show reproducible memory behavior.* **e.** *A comparison of the critical switching current density in 2D magnet-based SOT devices. The operating temperature of the devices is shown on the x-axis. Data points were obtained from references[28,30,167–173,31,158–161,163,165,166].* **f.** *State of the art of the field-free SOT-induced magnetic moment switching devices, spin Hall*

conductivity $\sigma_{SH}$ vs. dissipation power density for magnetization switching, P = ($J_{sw}^2/\sigma_c$). Data points were obtained from references [27,30,181–183,31,174–180].

Another exciting direction in the study of spin-orbit torques in 2D magnets is the emerging reports of **self-induced spin-orbit torques** (SSOT) in 2D ferromagnets, eliminating the traditional requirement of a spin source layer. Recent experimental works have revealed SSOT in $Fe_3GeTe_2$[184–186] and Co-doped $Fe_5GeTe_2$[149] and $Fe_3GaTe_2$[187–189]. In these materials, sizable damping-like and field-like torques arise intrinsically from the magnetic layer itself, in contrast to conventional spin Hall–driven torque in heavy-metal/ferromagnet heterostructures. Experimental signatures include current-induced coercivity modulation, harmonic Hall measurements, and direct observation of magnetization switching in all-vdW device geometries, even for very thick vdW magnet layers. The origin of the self-torque is attributed to strong spin–orbit coupling combined with broken inversion symmetry and large Berry curvature near the Fermi level, which enables efficient charge-to-spin conversion within the ferromagnet[190,191]. First-principles calculations and symmetry analysis suggest a close connection to intrinsic magnetic spin Hall and orbital Hall effects, with orbital angular momentum playing a key role in torque generation[192]. Compared to conventional SOT systems, SSOT exhibit exceptionally large torque efficiencies, enabling low-current-density switching and providing a pathway toward simplified, energy-efficient, all–van der Waals spintronic devices.

SSOT in vdW ferromagnets presents several unresolved challenges despite its strong potential for simplified spintronic devices. The microscopic origin of SSOT remains unclear, with competing mechanisms including intrinsic magnetic spin Hall effects, orbital Hall effects, and symmetry-breaking–induced spin accumulation not yet conclusively distinguished[193]. Experimentally, SSOT–driven switching often shows partial, nonuniform, or stochastic behavior and frequently requires an external magnetic field, indicating limited control over torque symmetry and magnetic anisotropy. In addition, SSOT has been observed in only a small number of metallic vdW magnets and is highly sensitive to crystal symmetry, thickness, stoichiometry, and disorder, raising concerns about reproducibility and scalability[149,184,190,191,193–195]. Finally, the thermal stability and long-term reliability of SSOT under high current densities remain largely unexplored, posing key challenges for its implementation in practical spintronic and neuromorphic devices.

**Future Perspective and Challenges**

2D magnet-based science and technology is at a **very early stage of research**, which needs the development of scalable materials platforms and integration with other materials and devices for practical applications. The technological impact of 2D spin-based devices is expected to evolve from relatively simple functionalities toward increasingly complex and integrated hardware platforms, as schematically summarized in **Fig. 6**. Presently, 2D magnet-based devices are mostly limited to exfoliated flakes from single crystals. It is very difficult to provide a timeline for these technological developments unless rapid progress is made in **scalable growth** techniques. Early spintronic applications, such as magnetic sensors and nonvolatile memory, rely primarily on well-controlled magnetic order, whereas more advanced paradigms, including spin logic, in-memory computing, neuromorphic architectures, and hybrid quantum spintronics, require precise control over interfacial spin interactions, spin transport, and nonequilibrium spin and orbital current induced magnetization dynamics[18,196,197]. 2D magnets are uniquely positioned along this roadmap, as their ultra-thin thickness, out-of-plane magnetic anisotropy, clean interfaces, and heterostructure compatibility enable systematic scaling of functionality and integration complexity within a unified materials platform.

A central opportunity offered by 2D materials is the ability to **engineer heterostructure** properties with high precision. Control over spin-orbit coupling, exchange interactions, and crystal symmetry can be achieved using external and internal parameters such as electrostatic gating, strain, twist angle, and interfacial proximity effects. These approaches allow systematic tuning of the proximitized electronic band structure of 2D heterostructures, charge-to-spin conversion efficiency in 2D spin-orbit materials, magnetic anisotropy, and exchange strength in 2D ferromagnets. Such control provides direct access to interfacial phenomena that are difficult to realize in conventional 3D systems and is expected to play a central role in optimizing spin-orbit torque efficiency and magnetization switching mechanisms.

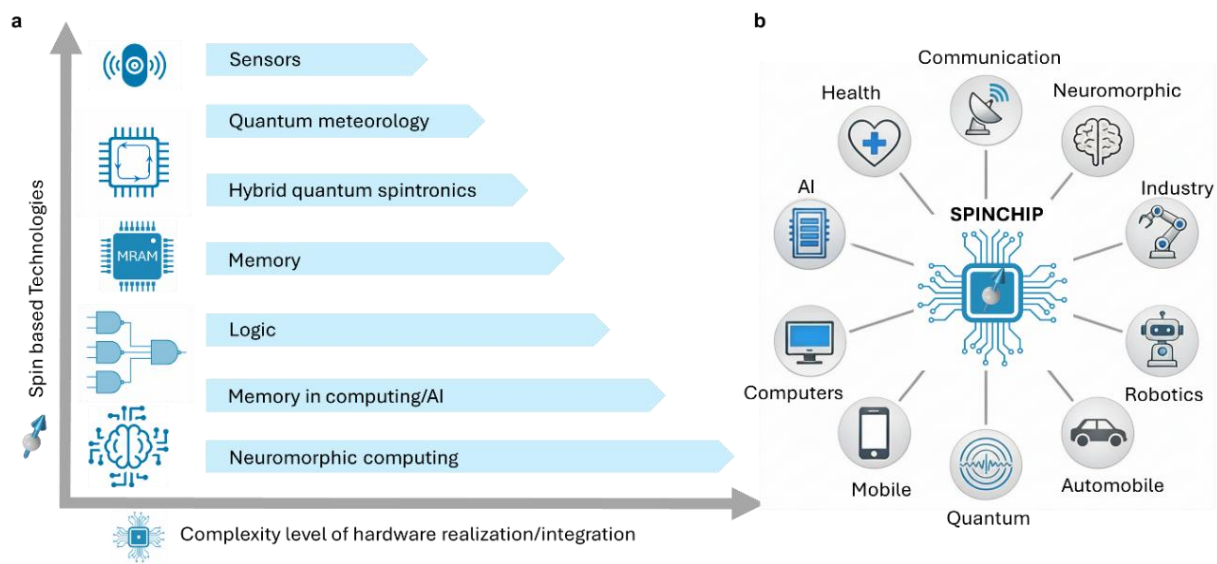

*Figure 6 | Roadmap of expected emerging 2D spintronic technologies. Conceptual technology roadmap for spin-based devices, illustrating the progression from relatively simple sensor and memory functionalities toward increasingly complex hardware platforms for quantum metrology, hybrid quantum spintronics, spin logic, in-memory computing, and neuromorphic architectures as integration complexity increases. As research and development on 2D magnets are at a very early stage, it is not possible to provide a timeline. Instead, we plotted the complexity levels of realizing these devices and circuits. b. Schematic illustration of a spin-based chip, denoted "SPINCHIP," at the center of an emerging technology ecosystem, highlighting prospective applications of spintronic and quantum spin devices in health, communication, neuromorphic computing, industrial automation, robotics, artificial intelligence, conventional computing, mobile electronics, quantum technologies, and automotive systems.*

From a device perspective, the diversity of magnetic ground states available in 2D materials enables application-specific design strategies. Ferromagnetic 2D systems are well-suited for nonvolatile memory and logic operations[198], while antiferromagnetic, ferrimagnetic and altermagnetic materials can offer faster magnetization dynamics, reduced stray fields, and intrinsic multistate behavior. Importantly, the partial, analog, and history-dependent switching characteristics observed in many SOT devices, often regarded as limitations for deterministic memory, become advantageous in emerging computing paradigms[199]. The atomically thin nature of 2D magnets makes them more susceptible to thermal fluctuations and external perturbations. As a result, stochastic and analog magnetization dynamics naturally emerge under electrical excitation, enabling multilevel, memristive responses mediated by domain nucleation and domain wall motion that closely resemble synaptic weight modulation[200]. When combined with SOT, voltage-controlled magnetic anisotropy, and proximity effects, 2D magnets can support probabilistic computing and ultra-low-energy synaptic operations, positioning them as promising platforms for **neuromorphic computing** beyond the conventional von Neumann architecture[199].

Beyond spin current-based magnetization switching, spin-wave or **magnon**-mediated signal transmission offer a promising route toward low-dissipation information processing and on-chip communication[201]. In insulating 2D magnets, particularly antiferromagnets and multiferroics, magnons can propagate over several micrometers, enabling energy-efficient spin-wave transport. 2D antiferromagnetic insulators are especially attractive in this context due to their high resonance frequencies, absence of stray fields, and robustness against external magnetic perturbations. When combined with electric-field control in multiferroic systems, spin-wave generation, propagation, and detection can be achieved with minimal power consumption, opening pathways for electrically reconfigurable magnonic circuits[202–204].

**Spin-wave** based communication complements charge-based spintronic memory and logic by acting as spin interconnects between functional blocks. Here, information is processed and stored locally using magnetic states, while magnons serve as carriers for inter-device communication, alleviating interconnect bottlenecks that limit conventional electronic architectures[205,206]. The atomic thickness and clean interfaces of 2D magnets are particularly advantageous for such hybrid architectures, as they enable strong coupling between magnonic, electronic, and ferroelectric degrees of freedom within the heterostructures. The combination of multistate magnetism, stochastic dynamics, and spin wave-based information transport further points toward computing paradigms **beyond von Neumann** architecture. Spin-based implementations of Heisenberg-type or analog computing, in which collective spin interactions encode and process information, offer an alternative route to machine intelligence that leverages intrinsic physical dynamics rather than Boolean logic[207]. 2D magnetic heterostructures provide a natural platform for these concepts, as exchange interactions, anisotropy, and coupling strength can be engineered through stacking, twist angle, strain, and electrostatic control.

Beyond classical spintronic applications, 2D magnets also open new directions in hybrid **quantum spintronic** architecture. The integration of 2D ferromagnets with superconductors has enabled the realization of ferromagnetic Josephson junctions and field-free Josephson diodes in vdW heterostructures[61]. In these systems, atomically sharp interfaces and symmetry breaking introduced by the magnetic barrier allow superconducting phase coherence to coexist with strong exchange interactions, overcoming the conventional incompatibility between magnetism and spin-singlet superconductivity. Experimental demonstrations of proximity-induced superconductivity in few-layer vdW ferromagnets, together with nonreciprocal Josephson transport in the absence of external magnetic fields, highlight the potential of these platforms for superconducting quantum circuits, cryogenic memory elements, and ultra-sensitive magnetometry[208–210]. More broadly, heterostructures combining magnets, superconductors, and spin-orbit materials provide a fertile platform for exploring unconventional superconductivity, phase-coherent spin transport, and electrically controllable quantum states. Hybrid quantum and sensing applications represent another promising frontier, where atomic-scale control, strong spin-orbit coupling, and long spin coherence may enable new approaches to quantum metrology and spin-based sensing.

Despite these exciting opportunities, several **challenges** must be addressed to enable the practical integration of 2D magnets into nanoelectronic technologies. A primary limitation remains the relatively low Curie temperature of most insulating and semiconducting 2D magnets, which often necessitates cryogenic operation. While metallic vdW ferromagnets such as $Fe_3GeTe_2$, $Fe_5GeTe_2$, Co or Ni-doped $Fe_5GeTe_2$ and $Fe_3GaTe_2$ have demonstrated near and above room temperature magnetic order, the experiments are so far limited to exfoliated flakes of several nanometer thickness. Getting ultra-thin layers of these materials in a reproducible manner still remains a challenge. For practical applications, achieving wafer-scale growth of such materials and reproducibility remains an outstanding materials challenge. To achieve a high Curie temperature above 400 K and address the issues, it will require atomic-level control over composition and structure through controlled alloying, chemical doping, strain engineering, and precise regulation of crystal symmetry and stoichiometry.

Equally important is the challenge of integrating high-quality 2D vdW **heterostructures** with reproducible interfaces. Variability arising from disorder, oxidation, and interfacial contamination can significantly affect magnetic anisotropy, spin transport, and torque efficiency. Developing scalable growth techniques for heterostructures, robust passivation strategies, and integration pathways compatible with existing semiconductor technologies will be essential for transitioning 2D magnetic systems from proof-of-concept demonstrations to reliable device platforms. Achieving deterministic control over magnetization dynamics while preserving the beneficial stochasticity required for neuromorphic and probabilistic computing will require careful co-design of materials and devices. Balancing thermal stability, switching energy, endurance, and variability will depend on a quantitative understanding of spin-orbit and orbital-mediated transport mechanisms. In this context, theory-guided materials discovery and symmetry-aware device engineering are expected to play a central role.

Overall, 2D magnetic materials provide a unifying framework that bridges nanoelectronics, spintronics, neuromorphic computing, and quantum technologies. As advances in materials synthesis, heterostructure engineering, and theoretical understanding continue, 2D magnets are poised to enable a new generation of compact, energy-efficient, and multifunctional nanoelectronic devices that exploit spin, charge, orbital, and quantum degrees of freedom within a single atomically engineered platform. Together, these developments suggest that 2D magnets may enable a fully integrated spin-based architecture in which memory, communication, and computation are co-designed at the level of fundamental magnetic interactions.

**Acknowledgments:** Authors acknowledge funding from European Innovation Council (EIC) project 2DSPIN-TECH (No. 101135853), European Innovation Council (EIC) project AMSWICH (No. 101258102), European Commission Graphene Flagship, 2D TECH VINNOVA competence center (No. 2019-00068), KAW – Wallenberg initiative on Materials Science for a Sustainable World (WISE), Swedish Research Council project (No. 2025-03702, and No. 2021-05925), FLAG-ERA projects 2DSOTECH (VR No. 2021-05925) and MagicTune, Graphene Center, AoA Nano, AoA Energy and AoA Materials program at Chalmers University of Technology.